\documentclass[aps,amsmath,twocolumn,amssymb,floatfix,superscriptaddress,footinbib]{revtex4-1}
\pdfoutput=1
\usepackage[dvips]{graphics}
\usepackage{bm}
\usepackage{epsfig}
\usepackage{enumerate}
\usepackage{subfigure}
\usepackage{color}
\usepackage{amsmath,amssymb,amsthm}
\usepackage{sidecap}
\usepackage{graphicx}
\usepackage{makecell}
\usepackage{hyperref}
\usepackage{multirow}
\usepackage{soul}

\usepackage[normalem]{ulem}

\hypersetup{
    pdfnewwindow=true,      
    colorlinks=true,       
    linkcolor=blue,          
    citecolor=magenta,        
    filecolor=blue,      
    urlcolor=blue        
}

\newcommand{\etal}{{\it et al.~}}
\newcommand{\ie}{{i.e. }}

\newcommand{\bea}{\begin{eqnarray}}
\newcommand{\eea}{\end{eqnarray}}
\newcommand{\beq}{\begin{equation}}  
\newcommand{\eeq}{\end{equation}}

\begin{document} 
\title{Finite bulk Josephson currents and chirality blockade removal from inter-orbital pairing in magnetic Weyl semimetals}

\author{Paramita Dutta}
\email{paramita.dutta@physics.uu.se}
\affiliation{Department of Physics and Astronomy, Uppsala University, Box 516, S-751 20 Uppsala, Sweden}
\author{Fariborz Parhizgar}
\email{fariborz.parhizgar@physics.uu.se}
\affiliation{Department of Physics and Astronomy, Uppsala University, Box 516, S-751 20 Uppsala, Sweden}
\author{Annica M. Black-Schaffer}
\email{annica.black-schaffer@physics.uu.se}
\affiliation{Department of Physics and Astronomy, Uppsala University, Box 516, S-751 20 Uppsala, Sweden}

\begin{abstract}
Magnetic Weyl semimetals (WSMs) have been presumed to be immune to proximity-induced spin-singlet superconducting pairing due to a chirality blockade of the regular Andreev reflection. In this work, we study all possible superconducting pairing induced in a WSM sandwiched between two conventional superconductors in a Josephson junction setup. We confirm that, while conventional  intra-orbital $s$-wave pairing is present on the surface of the WSM, it cannot propagate into the bulk due to the chirality blockade. However, inter-orbital $s$-wave pairing, with both even-frequency spin-singlet and odd-frequency mixed-spin-triplet symmetry, propagates into the bulk of the WSM, as do several $p$-wave symmetries. To demonstrate the importance of these finite inter-orbital and $p$-wave pair amplitudes in an experimental setup, we calculate the Josephson current and find a finite and even increasing current when the chirality blockade effect for the conventional intra-orbital pairing is enhanced.
\end{abstract}

\maketitle

\section{Introduction}
Weyl semimetals (WSMs) are one of the recently identified $3$D topological semimetals with isolated gapless and topologically protected Weyl nodes in the bulk energy spectrum, which are connected by unusual surface Fermi arcs\,\cite{Ashvin1,Burkov,AshvinRMP,raoweyl,yan2017topological}. The low-energy behavior near the Weyl nodes follow the Weyl Hamiltonian with fixed chiralities at different nodes. The number of Weyl points, always being even, depends on if the WSM is created by breaking the inversion or time-reversal symmetry, the latter producing a magnetic WSM. This unusual bulk band structure and its associated surface Fermi arc states have already been shown to give rise to many interesting properties, including chiral anomalies and the anomalous Hall effect \cite{AHEyang, BurkovChiralAnomaly, PRXChiralAnomaly, AshvinRMP}. Experimentally, several materials of the transition-metal pnictide family, such as TaAs and TaP, and others like NbAs have been recently realized to show signatures of the inversion-symmetry broken WSM phase\,\cite{xu2015discovery,xu2015Nature,PhysRevXTaAs,LeeTaP,PRXChiralAnomaly,xu2015experimental}, while GdPtBi\,\cite{hirschberger2016chiral}, Co$_3$Sn$_2$S$_2$\,\cite{Morali1286}, Co$_2$MnGa\,\cite{Belopolski1278}, and the kagome lattice\,\cite{Liu1282} have been identified as magnetic WSMs.

Not only is the normal state of WSM highly unusual, but adding superconductivity has been shown to produce exotic phases with bulk Bogoliubov-Weyl nodes\,\cite{balents,alidoust2019evolution}, chiral Majorana surface modes\,\cite{balents,hosur,yang,alidoust}, finite-momentum pairing\,\cite{cho,bednik2015}, and odd-parity superconductivity\,\cite{wei} in addition to $s$-wave superconductivity\,\cite{lu}. All these phenomena have so far only been considered in setups where superconductivity is intrinsic to the WSM, which, however, might be hard to realize experimentally, due to the semimetal nature of the WSM. On the other hand, proximity-induced superconductivity is a very attractive option for any advanced material. 

Proximity-induced superconductivity has already been studied in both inversion-symmetry-broken \cite{Franzproximity} and  magnetic \cite{KhannaProximity, Beenakker2017} WSMs using conventional $s$-wave spin-singlet superconductors (SCs). In particular, Bovenzi \etal identified the phenomenon of chirality blockade of the Andreev reflection between a magnetic WSM and a conventional $s$-wave SC\,\cite{Beenakker2017}, thus arguing for a complete lack of proximity effect from the bulk states of the WSM. 
The chirality blockade occurs because the backscattering of an electron as a hole, i.e.~Andreev reflection, must involve a switch of the Weyl node in order to preserve conventional spin-singlet pairing with its transfer of zero-spin and zero-momentum\,\cite{andreev1}. However, a change of the Weyl node in a magnetic WSM always results in a change in chirality and thus, the Andreev reflection process becomes forbidden. The natural conclusion is then that the WSM becomes immune to the superconducting proximity effect from a conventional superconductor, as the superconducting pairs cannot propagate through the bulk of the WSM. 

So far the only known intrinsic remedy for the chirality blockade in proximity-coupled magnetic WSMs is the surface Fermi arcs \cite{Beenakker2017, faraei2019}, as they do not suffer from the same chirality constraint and can thus still host proximity-induced spin-singlet $s$-wave pairing. However, with their exponential decay into the bulk of the WSM, their effect is vanishingly small in longer junctions. In addition, it has been shown that externally applying a magnetic Zeeman field or considering a spin-orbit active interface can also alleviate the chirality blockade \cite{Beenakker2017}.

Another way to mitigate the chirality blockade not yet investigated is the possibility that the WSM actually hosts superconductivity with other symmetries for the paired electrons. While the chirality blockade holds for spin-singlet $s$-wave pairing, the two-orbital nature and strong spin-orbit coupling in magnetic WSMs open up for multiple other alternatives, as then the Andreev reflection can also involve switching between the two orbitals or bands or spin-flips\,\cite{PhysRevB.81.224510,Niu_2012,PhysRevLett.97.067007,PDTI}. In particular, not only do the intrinsic two low-energy orbitals in magnetic WSMs generate the previously considered intra-orbital pairing, but also inter-orbital pairs can exist. In addition, spin-orbit coupling together with broken time-reversal symmetry not only allow for spin-triplet $p$-wave pairing, but they can also produce spin-triplet $s$-wave superconducting pairs with an odd time ordering, or equivalently an odd-frequency (odd-$\omega$) dependence \cite{Rmp,linder2017odd}. As a tangentially related example, odd-$\omega$ pairing has recently been shown to be the only surviving pair amplitude in proximity-coupled magnetic Weyl nodal loop semimetals due to their characteristic drumhead surface states \cite{parhizgarWNLS, duttaWNLS}.

In fact, the only general restriction on the symmetries of the superconducting pairing is a preserved fermionic nature of the Cooper pairs, leading to the so-called $\mathcal{SPOT} = -1$ classification of superconductivity \cite{AnnicaMulti, TriolaDriven, linder2017odd,triolaReviewMultiband}. Here $\cal{S}$ is the spin-symmetry, $\mathcal{P}$ the spatial parity, $\cal{O}$ the orbital parity, and $\mathcal{T}$ the symmetry in the relative time coordinate for the Cooper electron pair. Thus, for example, odd-$\omega$ pairing, first predicted by Berezinskii~\cite{BerezinskiiHe3}, can be in a spin-triplet configuration but still have $s$-wave intra-orbital spatial symmetry. Alternatively, actively involving the orbital index alternatively allows for e.g.~even-$\omega$ spin-triplet odd inter-orbital $s$-wave pairing. Particularly, odd-$\omega$ pairing can be generated by breaking any one of the spin, parity, orbital or time symmetries. In a proximity-coupled material it can even be induced using a conventional $s$-wave even-$\omega$ superconductor since the symmetry of the Cooper pairs may change according to the (lack of) symmetries in the proximity-coupled material.

In this work we consider a Josephson junction setup consisting of a magnetic WSM in-between two conventional $s$-wave spin-singlet SCs and we investigate in detail the proximity-induced pairing. We find that most allowed pair symmetries also exist in the WSM, this including intra- and inter-orbital, as well as even- and odd-$\omega$ pairing, with both $s$- and $p$-wave symmetries and in the spin-singlet and spin-triplet configurations. 
By increasing the magnetization to enforce a putative stronger chirality blockade, we indeed find that the intra-orbital $s$-wave spin-singlet, as well as the mixed-spin triplet odd-$\omega$ pairing, decrease quickly to zero inside the WSM, thus suffering from a chirality blockade. The intra-orbital $s$-wave equal-spin triplet pair amplitude is also vanishingly small within the WSM, resulting in no intra-orbital $s$-wave pairing remaining  in the interior of the WSM. However, inter-orbital $s$-wave pairing increases even when the chirality blockade prevents all intra-orbital pair amplitudes. In fact, in the middle of the WSM we find both large inter-orbital mixed-spin triplet odd-$\omega$ pairing and spin-singlet even-$\omega$ pairing. Beyond the isotropic and disorder robust $s$-wave states \cite{Michaeli&Fu12}, we find that also several $p$-wave amplitudes, with both intra- and inter-orbital configurations, survives within the WSM. 

As the superconducting proximity effect is driven by Andreev reflection at the WSM-SC interface, our results show that the chirality blockade is only present in some very specific pairing channels and that even a conventional Josephson junction allows for multiple other pair amplitudes in magnetic WSMs not suffering from any blockade. We verify the experimental importance of the finite superconducting pairing terms by showing that the Josephson current is large and even increasing when the magnetization is enhanced, despite the conventional intra-orbital pairing then being subjected to a complete chirality blockade. In the rest of the work we present these results in detail, starting with introducing our system and general model for a WSM Josephson junction in Section  \ref{mm}, then discussing all finite proximity-induced pair amplitudes in the WSM in Section \ref{pamp}, followed by calculating the critical Josephson current in Section \ref{JJ}, before we finally offer a few concluding remarks in Section \ref{conclu}.
 
\section{Model and Hamiltonian}
\label{mm}
We consider a generic model Hamiltonian for the magnetic WSM implemented on the cubic lattice as \cite{Beenakker2017}
\bea\label{Ham_wsm}
\bm{H}_{\text{\tiny WSM}} (\bm{k})&=&\tau_z \sum\limits_{i} t_i \sigma_i \sin {(k_i a)}
+  m(k) \tau_x \sigma_0+\beta \tau_0 \sigma_z  - \mu_{\text{\tiny W}}, \nonumber \\
\eea
with $m(k)=m_0+\sum\limits_i t^{\prime}_i (1-\cos{(k_i a)})$, and Pauli matrices $\tau_i$ and $\sigma_i$ acting in orbital and spin spaces, respectively. The two orbitals are $s$-like and $p$-like, thus having opposite parities which opens up for unusual odd inter-orbital pairing.
\begin{figure}[!thpb]
\centering
\includegraphics[height=2.9cm,width=0.99 \linewidth]{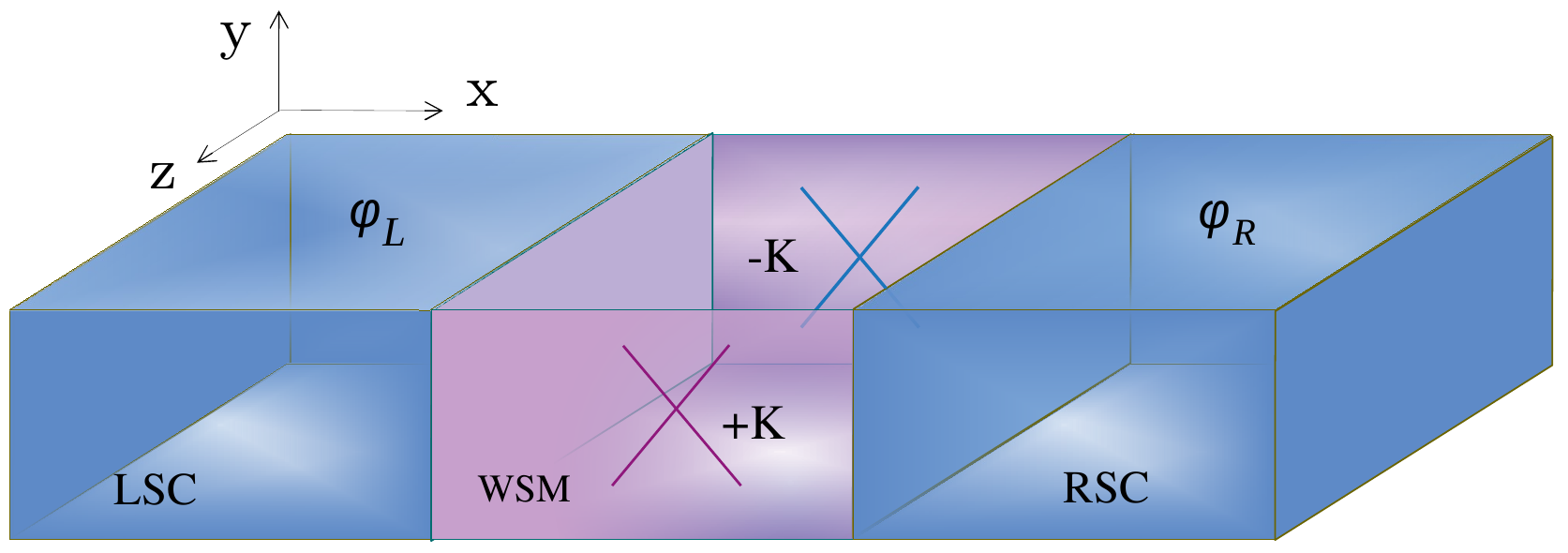}
\caption{Schematic diagram of Josephsoon junction set-up of a WSM (pink) positioned in-between two conventional SCs (blue) with superconducting phases $\varphi_{R,L}$. The dispersion drawn for the WSM is in reciprocal space and is just indicated for illustrative purposes.}
\label{model}
\end{figure}
Here, $k_i$ is the wave vector along the $i$-th ($x$, $y$ or $z$) axis, $a$ is the lattice constant, $t_i$ and $t'_i$ are nearest neighbor hopping parameters in the $i$-th direction, with the former being spin-dependent, $\beta$ is acting as a time-reversal symmetry breaking parameter or simply magnetization, and $\mu_{\text{\tiny W}}$ is the chemical potential. This generic model has two Weyl nodes located at ($0$,$0$, $\pm K$), with $K=\sqrt{\frac{\beta^2-m_0^2}{t_z^2+t_z^{\prime}m_0}}$\,\cite{Beenakker2017}. Thus, to create a WSM with two well-separated Weyl nodes, $\beta$ has to be larger than $m_0$, as $\beta=m_0$ describes a complete overlap of the two Weyl nodes. We also assume that $\beta$ is larger than the superconducting gap parameter to cover the experimentally relevant energy range\,\cite{Beenakker2017}. 
Furthermore, the direction of the separation of the Weyl nodes is set to be parallel to the WSM/SC interface in order to create the conditions for the chirality blockade and is also the same as in Ref.\,[\onlinecite{Beenakker2017}]. 

To model the whole Josephson junction, we assume a finite slab of the WSM tunnel-coupled to two conventional spin-singlet $s$-wave SCs,  left SC (LSC) and right SC (RSC), as shown schematically in Fig.\,\ref{model}. 
For the SCs, we assume the generic normal-state Hamiltonian
\bea
\bm{H}^{\text 0}_{\text{L(R)SC}}(\bm{k})=-\mu_{\text{SC}}+t_{\text{SC}}(2-\sum\limits_i\cos{(k_i a)}),
\label{hsc}
\eea
where $\mu_{\text{SC}}$ is the chemical potential of the SCs and $t_{\text{SC}}$ is the nearest neighbor hopping amplitude. 
To create the Josephson junction we discretize the two normal-state Hamiltonians, Eqs.\,(\ref{Ham_wsm}) and ({\ref{hsc}}), by taking the inverse Fourier transformation along the junction, the $x$-axis, while keeping translational invariance and thus the reciprocal space description along the $y$- and $z$-axes. 
We denote each layer of the WSM by $n_{\text w}$, which runs from $n_{\text w}=1$ to $n_{\text w}=N_{\text{\tiny WSM}}$, $N_{\text{\tiny WSM}}$ being the total number of layers in the WSM along the $x$-axis. Similarly, we include sufficient layers $N_{\text{SC}}$ for each SC. To connect the WSM and SCs we assume a spin-independent tunneling Hamiltonian $\bm{H}_{\text{T}}$ between neighboring layers of the SCs and WSM using the tunneling amplitude $t^{\tau}_{\text{\tiny{W-SC}}}$, with $\tau$ being the orbital (set to $a$ or $b$) index. We also denote the orbital tunneling ratio by $\tilde{t}=t^a_{\text{\tiny{W-SC}}}/t^b_{\text{\tiny{W-SC}}}$. For the exact expression of the tunneling Hamiltonian and additional details on the layer discretization, we refer to Appendix\,\ref{apn1}. 

After combining the WSM with the two SCs, the normal-state Hamiltonian for the whole LSC-WSM-RSC system has the form
\begin{align}
\bm{H}_{\text{W-SC}}(k_y,k_z) =\begin{pmatrix}
\bm{H}_{\text{LSC}}^{\text 0}&\bm{H}^{\dagger}_{\text T}&0\\
\bm{H}_{\text T}&\bm{H}_{\text {WSM}}&\bm{H}_{\text T}\\
0&\bm{H}^{\dagger}_{\text T}&\bm{H}_{\text{RSC}}^{\text 0},
\end{pmatrix}
\end{align}
where we have for notational simplicity suppressed the $k_y$ and $k_z$ dependences on the right hand side. 
Next, we introduce a Nambu basis and write the Bogoliubov-de Gennes (BdG) equation as
\beq
\bm{H}_{\rm BdG}(k_y,k_z) \Psi(r,k_y,k_z) = E \Psi(r,k_y,k_z),
\eeq
where $r$ corresponds to the site number along the $x$-axis, belonging to LSC or RSC or WSM. The BdG Hamiltonian for the whole LSC-WSM-RSC junction is
\beq
\bm{H}_{\rm{BdG}}(k_y,k_z)=
\begin{pmatrix}
\bm{H}_{\text{W-SC}}(k_y,k_z) & \bm{\Delta} \\
\bm{\Delta}^\dagger & -\bm{H}_{\text{W-SC}}^*(-k_y,-k_z)
\end{pmatrix},
\label{Hmat}
\eeq
when using the basis $\Psi=(\psi^{\dagger}_{r\uparrow}$,$\psi^{\dagger}_{r \downarrow}$,$\psi_{r \uparrow}$,$\psi_{r \downarrow})^T$, where again we suppress the $k_{y(z)}$ dependence. The gap matrix $\bm{\Delta}$ contains the superconducting order parameters and is expressed as 
\bea
\bm{\Delta}=\text{diag}( \Delta e^{i \varphi_{\text L}} , 0,\Delta e^{i \varphi_{\text R}} )   \otimes\tau_0 \otimes \sigma_y,
\eea 
where diag represents a diagonal matrix in the spatial index $r$ throughout the whole Josephson junction, with $\Delta$ being the magnitude of the superconducting order parameter in the two SCs (also the energy gap) and $\varphi_{\text{L(R)}}$ the phase of the L(R)SC.

Throughout this work, we fix the energy parameters to $t^{\prime}_i=t_i=t_{\text{SC}}=1$, $m_0=0.1$, and the lattice constant to $a=1$. Moreover, we consider mainly undoped WSMs with $\mu_{\text W}=0$, unless explicitly mentioning another doping (inset of Fig.\,\ref{figJ}). For the SCs we set $\mu_{\text SC} = 2$ and $\Delta=0.1$. We use zero phase $\varphi_L=\varphi_R=0$ when studying the pair amplitudes, but set the phase difference to $\varphi_R-\varphi_L=\pi/2$ to find the maximum Josephson current (appearing very close to $\pi/2$). 
For the WSM-SM tunneling process, we set the two orbital tunneling amplitudes equal, $t^{a}_{\text{\tiny {W-SC}}}=t^b_{\text{\tiny {W-SC}}}=0.5<t$, except in Fig.\,\ref{figt}, where we study asymmetric tunneling. We show results for $N_{\text {\tiny WSM}}=41$ layers of the WSM and $N_{\text {SC}}=40$ SC layers, but have also studied other system sizes. This system size choice is sufficient to induce bulk conditions in both the interior of the WSM and the SC leads as the SC coherence length, $\xi =\hbar v_{Fx}/\Delta \sim 10 a$ (using the Fermi velocity along the $x$-direction), is much smaller than the system size. All parameters have realistic values, except the SC gaps that are a bit larger in order to get a finite Josephson current for a still computationally manageable system size. However, our assumptions for all parameter values do not affect the qualitative behavior of our numerical results.

\section{pair amplitude}\label{pamp}
In order to investigate proximity-induced superconductivity in the WSM and its symmetries, we 
compute in this section the anomalous Green's function for the whole Josephson junction and discuss its behavior. The anomalous Green's function captures the pair propagation in a system and therefore becomes a direct measure of the proximity-induced pair amplitude in the WSM. To extract the anomalous Green's function, we start with the definition of the retarded Green's function given by
\bea
\bm{G}^R(\omega,k_y,k_z)=[(\omega+i \delta) \bm{I}-\bm{H}_{\rm BdG}(k_y,k_z)]^{-1}, 
\eea
which we express in a block-form in the Nambu basis:
\beq
\bm{G}^R(\omega,k_y,k_z)=
\begin{pmatrix}
\bm{\mathcal{G}}(\omega,k_y,k_z) & \bm{\mathcal{F}}(\omega,k_y,k_z) \\
\bar{\bm{\mathcal{F}}}(\omega,k_y,k_z) & \bar{\bm{\mathcal{G}}}(\omega,k_y,k_z)\end{pmatrix}.
\label{Gmat}
\eeq
From this we have direct access to the anomalous part of the Green's function, $\bm{\mathcal{F}}(\omega,k_y,k_z)$ which is the pair amplitude $\langle T \Psi(\omega) \Psi (0)\rangle$, with $T$ being the time-ordering operator\,\cite{BalatskyMeissner}. Here ${\mathcal{F}}$ is an $N$x$N$-dimensional block matrix, with $N=2(2N_{\text{\tiny WSM}}+2N_{\text{\tiny SC}})$. Focusing on the WSM, for each layer $r$ in the WSM it takes the form
\bea
\mathcal{F}_{\tau\,\tau^{\prime}}(\omega)&=&\sum\limits_{k_y,k_z}{\mathcal{F}}(\omega,k_y,k_z)\nonumber \\
&=&
\begin{pmatrix}
\mathcal{F}_{\tau\uparrow\; \tau^{\prime}\uparrow}(\omega) & \mathcal{F}_{\tau \uparrow\, \tau^{\prime}\downarrow}  (\omega)\\
\mathcal{F}_{\tau\downarrow\, \tau^{\prime} \uparrow}(\omega) & \mathcal{F}_{\tau\downarrow\, \tau^{\prime} \downarrow}(\omega)
\end{pmatrix},
\label{Gehmat}
\eea
which makes the intra-orbital ($\mathcal{F}_{\text{aa}}$) and inter-orbital ($\mathcal{F}_{\text{ab}}$) nature evident. Here the diagonal components provide the equal spin-triplet components ($\uparrow \uparrow$ and ${\downarrow \downarrow}$), while the off-diagonal elements combine into the mixed-spin triplet ($\uparrow \downarrow$$+$$\downarrow \uparrow$) and spin-singlet ($\uparrow \downarrow$$-$${\downarrow \uparrow}$) components. 

In our system, if we consider a summation of $k_y$ and $k_z$ over the whole Brillouin zone, both the full bulk and surface Fermi arc contributions are automatically taken into account. Now, the Fermi arc states are known to violate the chirality blockade since they are not part of the Weyl spectrum, as discussed in Ref.~[\onlinecite{faraei2019}]. On the contrary, our main focus is the contribution by the bulk states to the proximity effect. Thus, in order to minimize the effect of the surface Fermi arcs, we limit the summation in the anomalous Green function over a region of width $2k_c$ ($k_c=\pi/8$) around each Weyl node so that we mainly capture the linear regime of the two Weyl cones but minimize the overlapping region which contain the Fermi surface arcs. With this procedure we still have a small contribution by the surface states but, as we show, it does not influence our results. Most importantly, the value of $k_c$ gives a corresponding energy in the Weyl node that is much larger than $\Delta$, such that we still cover a broad energy range, including all the Andreev states within the superconducting gap. We verify this by taking the summation over the whole Brillouin zone for which we recover intra-orbital pairing also for larger $\beta$, in addition to the significant inter-orbital pairing we report also without the Fermi arc contributions. The value of $k_c=\pi/8$ is a particular choice here, but we find that other choices for $k_c$ fulfilling the above conditions also work.

\subsection{$s$-wave pair amplitudes}
We start our discussion of the proximity-induced pairing in the WSM by considering pair amplitudes with isotropic $s$-wave spatial symmetry. Due to its lack of $k$-dependence, this pair amplitude is stable even in the presence of disorder and is thus usually the most relevant pair symmetry. We note that the disorder robustness extends also to inter-orbital $s$-wave pair amplitudes even if they are odd under orbital parity, due to the large spin-orbit coupling in WSMs \cite{Michaeli&Fu12}.

\begin{figure}[htb]
\includegraphics[width=\columnwidth]{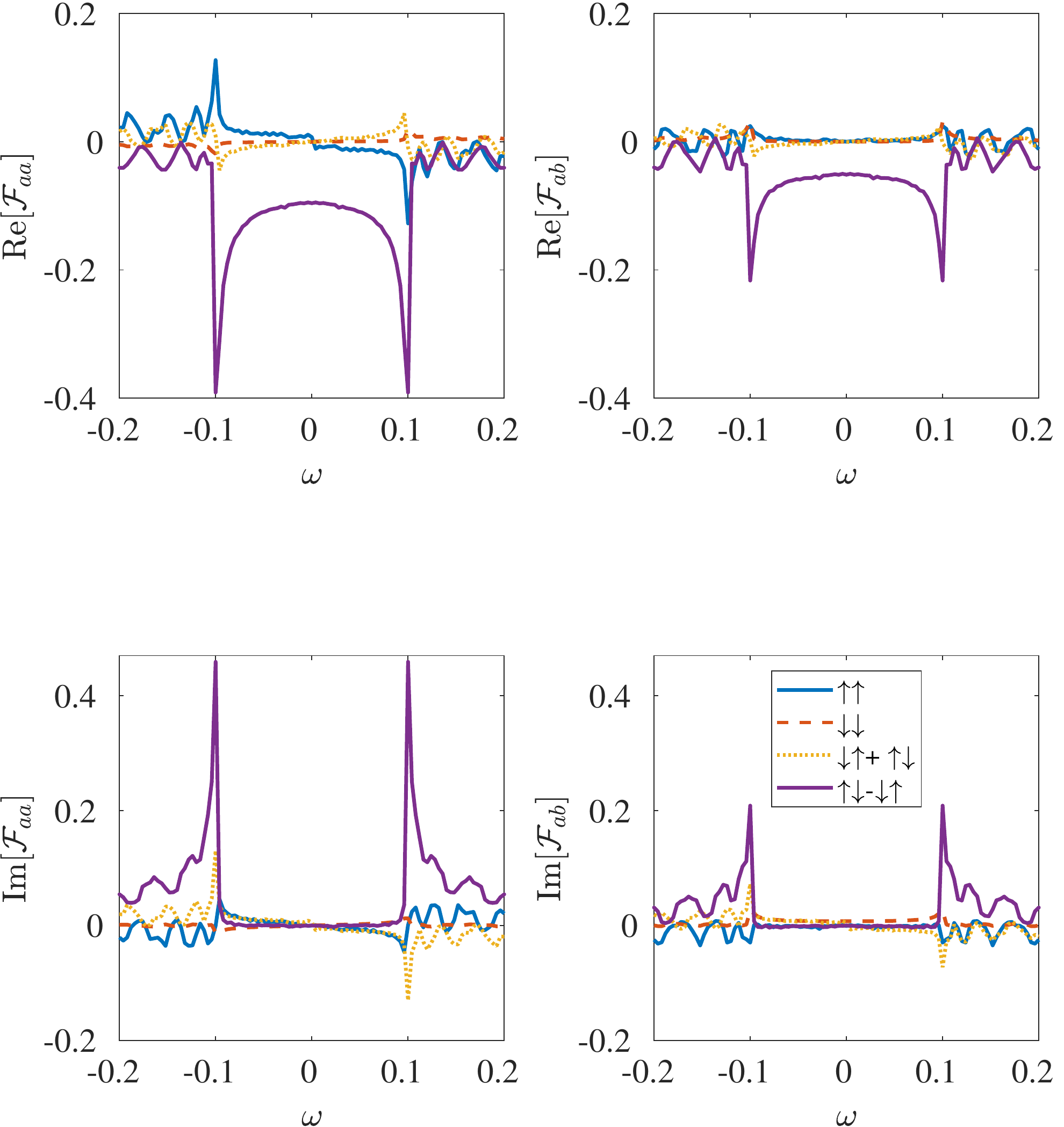}
\caption{Real (top) and imaginary (bottom) parts of the $s$-wave intra- (aa) and inter-(ab) orbital pair amplitudes as a function of frequency in the first layer of the WSM ($n_{\text w}=1$) for $\beta=0.2$.}
\label{figw}
\end{figure}

We extract all isotropic $s$-wave pair amplitudes by allowing all possible configurations in spin and orbital space while keeping the real space extent to on-site pair amplitudes.
In Fig.\,\ref{figw} we show the real and the imaginary parts of the anomalous Green's function as a function of frequency $\omega$ in the first ($n_{\text w}$$=$$1$) layer of the WSM for a specific, and rather low, $\beta=0.2$. We plot separately all the four different spin configurations: spin-singlet ($\uparrow\downarrow$$-$$\downarrow\uparrow$), equal-spin triplet ($\uparrow \uparrow$ and $\downarrow\downarrow$) and mixed-spin triplet ($\uparrow\downarrow$$+$$\downarrow\uparrow$) pairing, while the left and right columns correspond to the intra-orbital and inter-orbital amplitudes, respectively. 
From both the real and the imaginary parts of the intra-orbital pair amplitude, we see directly that all the three spin-triplet amplitudes are odd functions of frequency, whereas the spin-singlet amplitude is an even function of frequency. This is fully consistent with Fermi-Dirac statistics for the Cooper pairs, since spin-triplet $s$-wave pairing acquires the necessary oddness under exchange of all quantum indices between the two electrons only for an odd frequency dependence, i.e.~an oddness in the relative time coordinate between the two electrons.
We note here that the spin-triplet amplitudes appear in the WSM even with conventional spin-singlet SC leads and no  spin-active interfaces. This is due to the internal spin-orbit coupling and magnetization in the WSM, which are sufficient conditions for the appearance of the spin-triplet pairing \cite{linder2017odd,parhizgarWNLS, duttaWNLS,JPSJTanaka}. 

Next, turning to inter-orbital pairing, we find that the equal-spin triplet, both $\uparrow\uparrow$ and $\downarrow\downarrow$ amplitudes, and the spin-singlet pair amplitude are all even functions of frequency, while the inter-orbital mixed-spin triplet has an odd frequency dependence. To verify that these pair amplitudes also satisfy the Fermi-Dirac statistics for Cooper pairs, we explicitly check the symmetry with respect to a change in the orbital index. 
We find that the inter-orbital equal spin-triplet even-$\omega$ amplitude is odd under the exchange of the orbital index, while the mixed-spin triplet pairing is an even function under the exchange of the orbital index, which then correctly, forces an odd-$\omega$ dependence for the latter. For the spin-singlet amplitude we find that it is even under the exchange of the orbital index and therefore it is also even in frequency. Thus all symmetries indeed satisfy the $\mathcal{SPOT} = -1$ classification of superconductivity \cite{AnnicaMulti, TriolaDriven, linder2017odd,triolaReviewMultiband}. While Fig.\,\ref{figw} only report the pairing in the first WSM layer (\ie closest to the SC), we have verified that these symmetries of the pair amplitudes  persist through the whole WSM lattice structure.

\begin{figure*}[htb]
\includegraphics[scale=0.43]{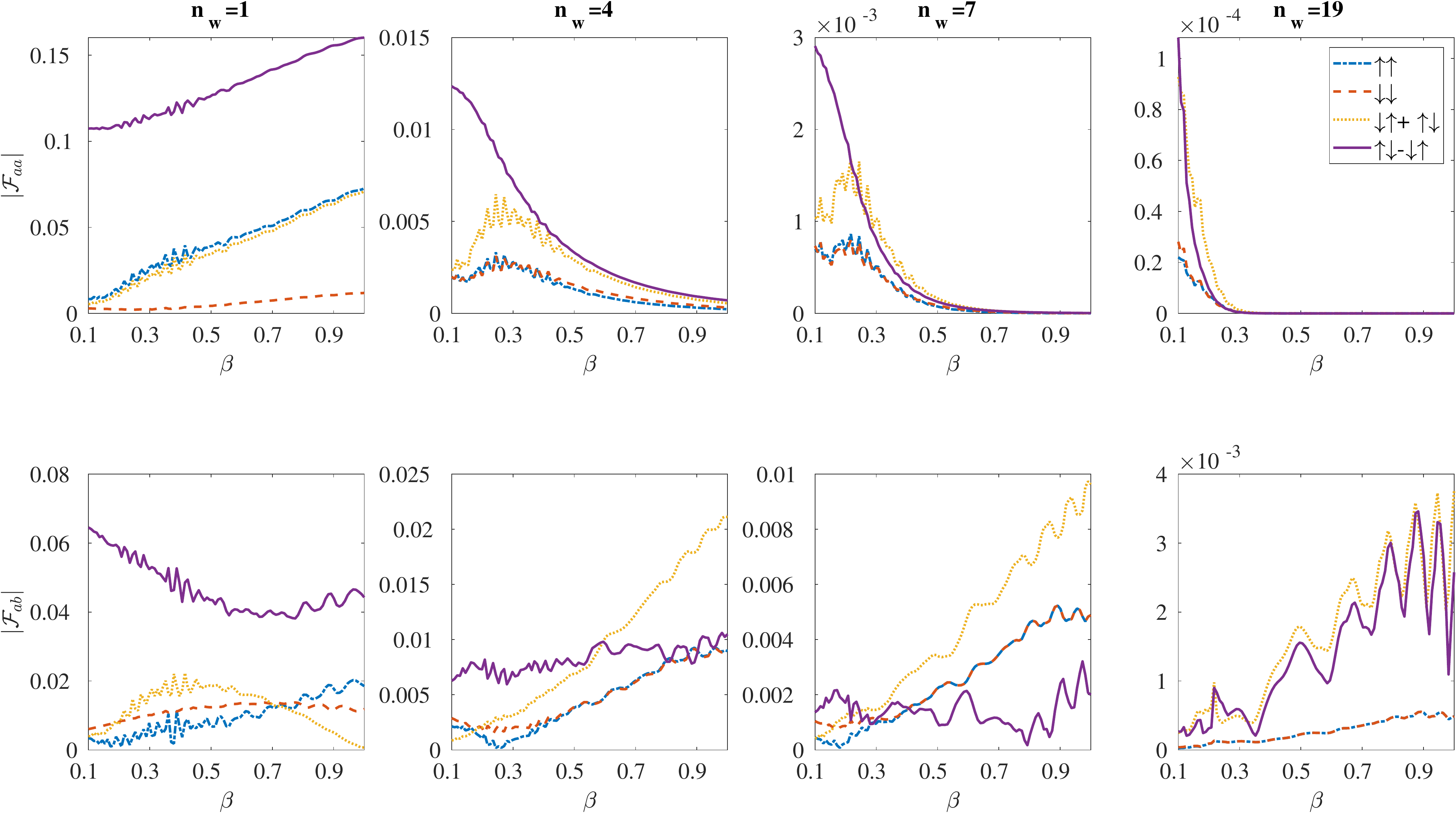}
\caption{Intra-orbital ($\mathcal{F}_{\text{aa}})$ and inter-orbital ($\mathcal{F}_{\text{ab}}$) pair amplitudes as a function of $\beta$ for different layers ($n_{\text w}$) of the WSM for $\omega=0.5\,\Delta$. Note the difference in the y-axis scales.}
\label{figBeta}
\end{figure*}

After having investigated the symmetries of all proximity-induced $s$-wave pair amplitudes in the WSM Josephson junction we next turn our attention to our primary topic, namely the strength of the chirality blockade. To investigate the chirality blockade we study the effect of the magnetization, given in terms of $\beta$, where larger $\beta$ generates a larger separation of the two Weyl cones, and thus presumably a larger chirality blockade effect.
In Fig.\,\ref{figBeta} we plot all pair amplitude as a function of $\beta$ for different layers of the WSM, where we set the lower limit of $\beta$ as $0.1$, the value of $m_0$. The upper and the lower panels represent the intra-orbital and inter-orbital pairing, respectively, with each column moving further into the WSM, eventually reaching bulk conditions. 

For the intra-orbital pairing in the upper panel, we observe that in the boundary layer of the WSM ($n_{\text w}=1$), spin-singlet pairing dominates over all the spin-triplet terms, similar to Fig.\,\ref{figw} which shows the result for a particular value of $\beta$. This behavior is expected as the attached SC leads have an order parameter with conventional $s$-wave spin-singlet symmetry. The ratio of the even-$\omega$ spin-singlet to the odd-$\omega$ spin-triplet pair amplitudes does not change considerably with the increase of $\beta$, although the more magnetization we apply, the higher is the proximity-induced pair amplitude. This is also expected since in the WSM surface layer, mostly the surface states contribute and an increase of $\beta$ generates a longer Fermi surface arc.
However, this scenario changes drastically when we move into the WSM. We choose to plot the pair amplitudes in layers $n_{\text w}=4$, $7$ and $19$ to display the gradual change in the behavior towards the middle layer of the WSM. Already at $n_{\text w}=4$ the amplitude of the spin-singlet amplitude decays with the increase of $\beta$, completely different from the behavior in the first layer. All intra-orbital triplet odd-$\omega$ pair amplitudes increase initially for small $\beta$ but eventually also decay. Moving even further into the WSM we find that this initial increase for small $\beta$ vanishes, and all amplitudes now just monotonically decay with increasing $\beta$. 

As we move into the bulk of the WSM, the distance from the SC leads increases and therefore the proximity-induced pairing must decrease, following a well-known exponential decay. However, on top of that, when we increase $\beta$, we find that all intra-orbital pair amplitudes decay exponentially to zero, with a decay rate that is largest well within the WSM.
This means that for sufficiently high magnetization $\beta$, it is not possible to proximity-induce intra-orbital superconducting pairing in the WSM, while for small magnetizations there is no such blockade and the proximity effect is more normal in nature. 
This can be understood by noting that increasing $\beta$ leads to an increased separation of the two Weyl nodes and thus a decreasing overlap between the two cones, which can host sub gap levels. As a consequence, increasing $\beta$  corresponds to an ever increasing proportion of the pairing between electrons within each Weyl cone. At the same time, proximity-induced pairing is associated with the phenomenon of Andreev reflection of the electrons at the interface\,\cite{andreev1}. Thus, for spin-singlet pairing to maintain a zero momentum and zero spin-transfer, the two electrons in the Cooper pair have to belong to the two opposite Weyl nodes of the WSM. But these two electrons then necessarily require opposite chirality and the Andreev reflection becomes blocked. This result exactly corroborates the finding of a chirality blockade in Ref.\,[\onlinecite{Beenakker2017}]. 

Similar to the intra-orbital spin-singlet pairing, the intra-orbital mixed-spin triplet odd-$\omega$ pair amplitude also vanishes with the spin-singlet pairing following the same reasoning with chirality blockade. However, for the intra-orbital equal-spin triplet pairing the Andreev reflection process transfers a finite spin moment and thus there is no inherent chirality blockade in the WSM for this pairing. Still, we only find equal-spin triplet pairing well within the WSM for $\beta \lesssim 0.2$. We instead find that the equal-spin triplet amplitude is suppressed by another mechanism: The surface Fermi arcs of the WSM are polarized with opposite spin orientations on the two surfaces, \ie at the two different SC leads, and also within the WSM there is a spin-symmetry such that the $\uparrow\uparrow$ triplet pair amplitude in layer $n_{\text w}=1$ is the same as that in layer $n_{\text w}=N_{\text{\tiny WSM}}$ but with opposite sign. This leads to the equal-spin triplet pair amplitude being exactly zero in the middle of the WSM and thus also heavily suppressed around this layer. The equivalent is true also for the $\downarrow\downarrow$ pair amplitude.
For larger $\beta$ this spin polarization phenomenon is more pronounced and thus suppresses the equal-spin triplet pair amplitudes even more effectively within the WSM. 

Having verified the existence of the chirality blockade of the intra-orbital spin-singlet and mixed-spin triplet pairing, and having shown how large magnetization also strongly suppresses the equal-spin triplet infraorbital state, we next turn our attention to the inter-orbital pairing, presented in the lower panesl of Fig.\,\ref{figBeta}. 
Similar to the intra-orbital case, spin-singlet symmetry dominates in the surface layer of the WSM, although the pair amplitudes do not generally increase with increasing $\beta$ on the surface. However, in the interior of the WSM, we observe that many pair amplitudes actually increases with $\beta$. This is in complete contrast to the intra-orbital amplitudes, which are all heavily suppressed by increasing the magnetization through $\beta$. 
In particular, we find that both the inter-orbital even-$\omega$ spin-singlet and odd-$\omega$ mixed-spin triplet states thrive within the WSM with a clear increase with increasing $\beta$. There is thus {\it no} chirality blockade for either of these inter-orbital pair amplitudes, but they remain large even far from the SC leads in also heavily magnetic WSMs. It is the addition of the orbital index that makes it possible to induce inter-orbital pairing in the WSM, as chirality blockade of the Andreev reflection is not necessarily present when pairing occurs between two electrons from different orbitals: electrons of opposite spins but from different orbitals may couple together without any change in the chirality. The exact structure of the orbital dependence depends on the type of the WSM, but since even our generic and simple WSM model captures a lack of chirality blockade for inter-orbital pairing, it offers an excellent counter argument to a generic chirality blockade of proximity effect in magnetic WSMs. We also note that the equal-spin triplet components, while zero in the exact middle of the WSM (checked for $n_{\text {\tiny W}}=21$), are generally growing with increasing $\beta$ in the interior of the WSM, which reinforces the idea  that proximity-induced pairing is present in magnetic WSM.

In Fig.\,\ref{figBeta} we choose a particular value of $\omega$ ($=0.5 \Delta$), which is well within the superconducting energy gap, but this choice of $\omega$ does not affect the qualitative behavior of the pair amplitudes, including choosing $\omega$ to be of the size of the superconducting gap. Also, all the results for the pair amplitudes are symmetric with respect to the middle layer of the WSM, \ie the results for all layers on the left side of the junction are exactly the same as those for the right side layers. In Table~\ref{tab} we summarize our results by providing a simple overview of which pair amplitudes are generally present.
\begin{table*}
\setlength{\tabcolsep}{10pt}
\renewcommand{\arraystretch}{1.5}
\begin{tabular}{|c|c|c|c|c|}
\hline
\bf{Parity} & \bf {Orbital}  & \bf{Frequency} & \bf{Spin} & \bf{Pair amplitude }\\
\hline \hline
 \multirow{6.5}{*}{$s$-wave} &\multirow{3.5}{*}{intra} & even &  singlet ($\uparrow\downarrow-\downarrow\uparrow$)& blocked by chirality\\  \cline{3-5}
&  &  \multirow{2.25}{*}{odd} & equal triplet ($\uparrow\uparrow$, $\downarrow\downarrow$) & \makecell{almost zero$^*$}\\  \cline{4-5}
&  & & mixed triplet ($\uparrow\downarrow+\downarrow\uparrow$)& blocked by chirality\\  \cline{2-5}
&  \multirow{3.5}{*}{inter}&  \multirow{2.25}{*}{even} & singlet ($\uparrow\downarrow-\downarrow\uparrow$) & higher by $2$x magnitude\\  \cline{4-5}
& & & equal triplet ($\downarrow\downarrow$, $\uparrow\uparrow$)& \makecell{higher by $1$x magnitude$^*$} \\  \cline{3-5}
&  & odd & mixed triplet ($\uparrow\downarrow+\downarrow\uparrow$)& higher by $2$x magnitude\\
\hline \hline
\multirow{6}{*}{$p_x$ and $p_z$-wave} & \multirow{3}{*}{intra} & \multirow{2}{*}{even} & equal triplet ($\uparrow\uparrow$, $\downarrow\downarrow$)& lower by 2x magnitude \\ \cline{4-5} 
& & & mixed triplet ($\uparrow\downarrow+\downarrow\uparrow$)& lower by 2x magnitude \\ \cline{3-5} 
&  & odd & singlet ($\uparrow\downarrow-\downarrow\uparrow$) & lower by 3x magnitude \\ \cline{2-5}
&  \multirow{3}{*}{inter} & even &mixed triplet ($\uparrow\downarrow+\downarrow\uparrow$)& lower by 1-2x magnitude$^*$ \\ \cline{3-5}
&  &  \multirow{2}{*}{odd}& singlet  ($\uparrow\downarrow-\downarrow\uparrow$) & lower by 1-2x magnitude$^*$ \\  \cline{4-5}
&  & & equal triplet  ($\uparrow\uparrow$, $\downarrow\downarrow$)& lower by 1-2x magnitude \\
\hline \hline
\multirow{6}{*}{$p_y$-wave} &  \multirow{3}{*}{intra}&  \multirow{2}{*}{even} & equal triplet ($\uparrow\uparrow$, $\downarrow\downarrow$)& lower by 1x magnitude$^*$\\ \cline{4-5}
& & & mixed triplet($\uparrow\downarrow+\downarrow\uparrow$) & higher by $2$x magnitude \\ \cline{3-5}
& & odd &singlet($\uparrow\downarrow-\downarrow\uparrow$) & higher by $2$x magnitude \\ \cline{2-5}
& \multirow{3}{*}{inter} & \multirow{2}{*}{even}& singlet  ($\uparrow\downarrow-\downarrow\uparrow$)& comparable \\  \cline{4-5}
& & & equal triplet ($\uparrow\uparrow$, $\downarrow\downarrow$)&lower by 1x magnitude$^*$ \\ \cline{3-5}
& & odd&  mixed triplet  ($\uparrow\downarrow+\downarrow\uparrow$)& comparable\\
\hline \hline
\end{tabular}
\caption{ \label{tab}Symmetry classification of all pair amplitudes in the interior of the WSM with their magnitudes given in the chirality blockade regime ($\beta \ge 0.5$). Comments on the $s$-wave pair magnitudes are made with respect to that of the even-frequency $s$-wave intra-orbital pairing, whereas comments for the $p$-wave pair amplitudes are made in comparison to the inter-orbital $s$-wave pair amplitudes with the same spin structure. Asterix ($^*$) refers to pair amplitude that is identically zero in the middlemost layer of the WSM.}
\end{table*}

\begin{figure*}[htb]
\includegraphics[scale=0.43]{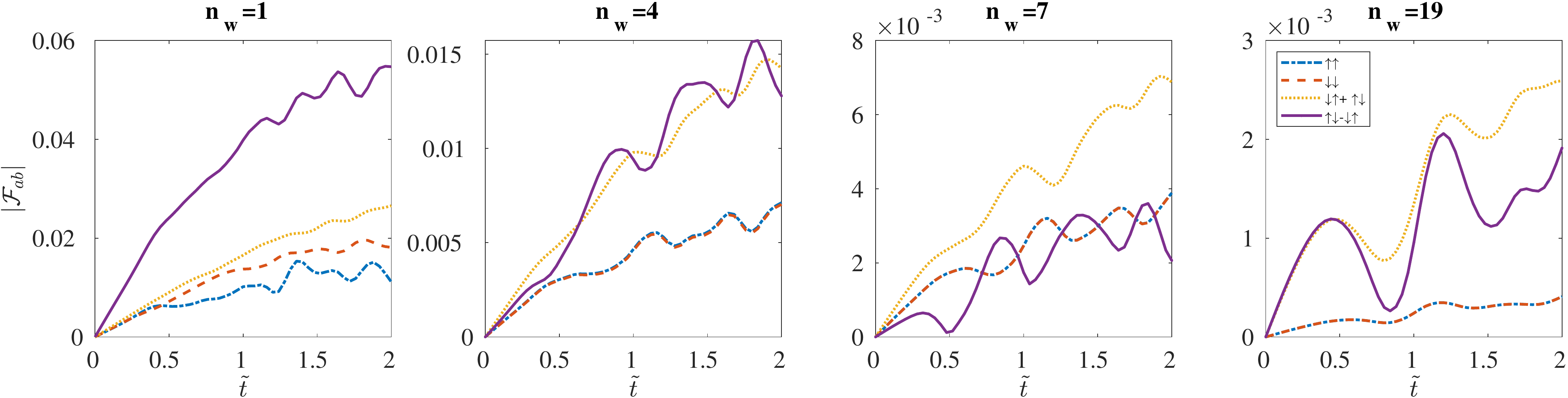}
\caption{Inter-orbital pair amplitude $\mathcal{F}_{ab}$ as a function of the WSM-SM tunneling asymmetry $\tilde{t}=t^a_{\text{w-sc}}/t^b_{\text{w-sc}}$ for $t^b_{\text{w-sc}}=0.5$ and $\beta=0.6$. Other parameter values are same as in Fig.\,\ref{figBeta}}.
\label{figt}
\end{figure*}

Up until now we have only considered the proximity effect to be equally large into both orbitals in the WSM. However, an asymmetry is likely present in real WSMs as the orbitals can easily be located at different distances from the SC-WSM interface. Considering that we have just shown that inter-orbital pairing plays a key role and that such coupling asymmetry have previously been shown to be important in other topological systems\,\cite{AnnicaTI,DushkoSH,FPbilayer}, we next investigate the behavior of the pair amplitudes as a function of tunneling asymmetry, measured by $\tilde{t}$, in Fig.\,\ref{figt}. Here we fix $t^b_{\text{w-sc}}=0.5$ and then vary $t^a_{\text{w-sc}}$ from 0 to $1$, producing the symmetric coupling situation for  $\tilde{t} = t^a_{\text{w-sc}}/t^b_{\text{w-sc}} =1$ and also an asymmetry between the $\tilde{t}>1$ and $\tilde{t}<1$ regimes. This will then break the inversion-symmetry locally. Further, we choose $\beta=0.6$ which corresponds to the regime of chirality blockade for the intra-orbital pairing, and therefore we only show the inter-orbital pair amplitudes. We display the results for the same four layers $n_{\text w}=1$, $4$, $7$ and $19$ of the WSM as in Fig.\,\ref{figBeta}. We find that, in general, all pair amplitudes increase with the rise of $\tilde{t}$. There are also oscillations at higher tunneling ratios, which extends to lower $\tilde{t}$ when moving into the interior of the WSM. In the middle WSM layer we again only have two finite pair amplitudes, true for all values of $\tilde{t}$, which therefore further supports the generality of the statement that finite inter-orbital pairing always exists in a magnetic WSM. 

\subsection{$p$-wave pair amplitudes}
\begin{figure*}[htb]
\includegraphics[scale=0.43]{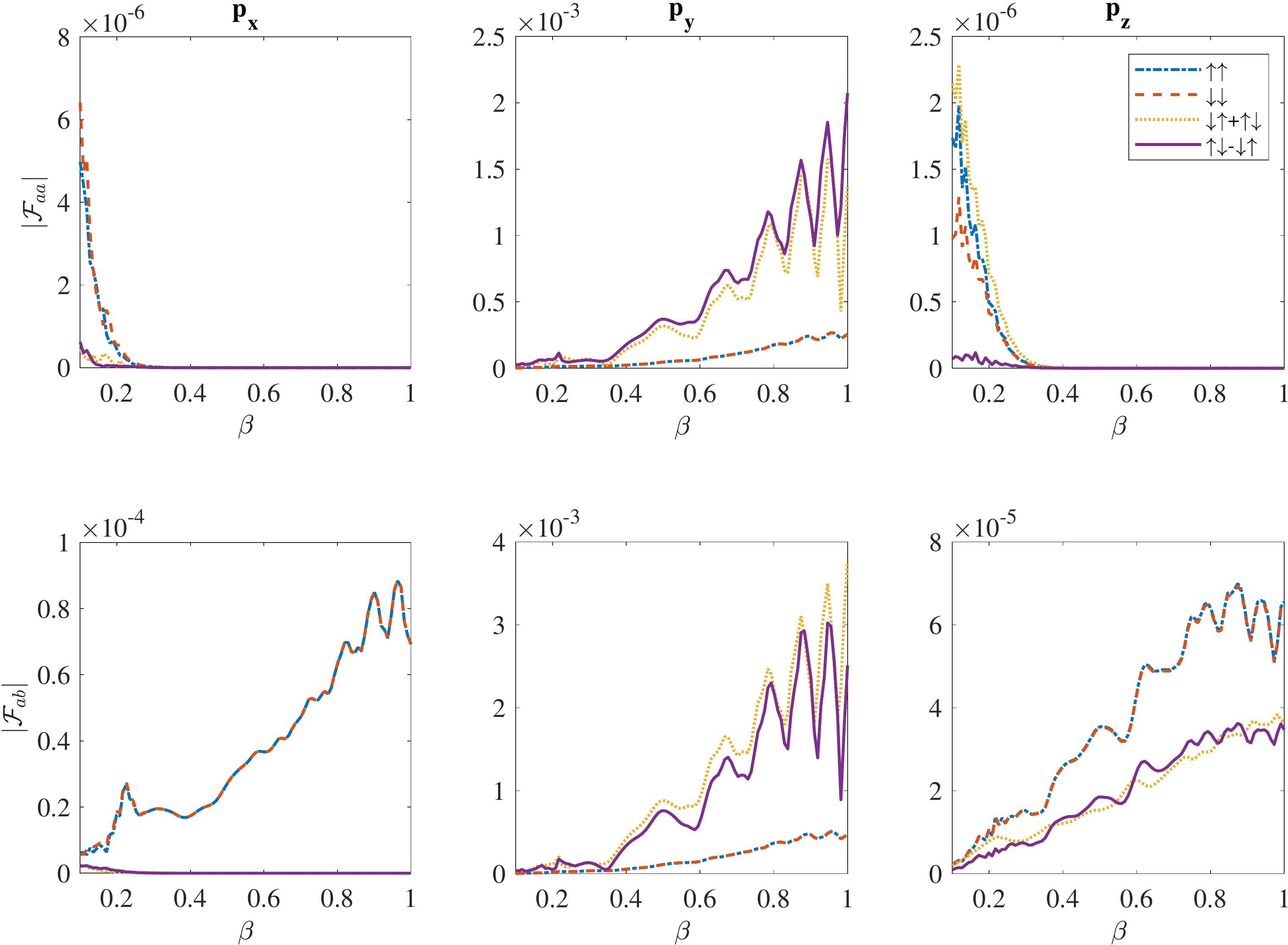}
\caption{Intra-orbital ($\mathcal{F}_{aa}$) and inter-orbital ($\mathcal{F}_{ab}$) pair amplitude of $p_x$, $p_y$ and $p_z$-wave symmetry as a function of $\beta$ in one of the bulk layers, $n_{\text w} = 19$. Other parameter values are the same as in Fig.\,\ref{figBeta}.}
\label{figp}
\end{figure*}

Having established the presence of inter-orbital $s$-wave pairing in magnetic WSM, independent of both the degree of magnetization and asymmetry in the WSM-SM tunnel coupling, we next turn to possible non-$s$-wave pair amplitudes. Of these, the $p$-wave amplitudes are most interesting, since they have the opposite spatial parity. However, due to non-$s$-wave symmetries not being isotropic in reciprocal space we do not expect them to play a major role in systems where disorder is present. Still, for the completeness of this study, we include the behavior of $p$-wave pair amplitudes in Table~\ref{tab}. 
In particular, we consider pairing between two nearest neighbor sites along the three different directions, $x$, $y$ and $z$, and we plot the corresponding $p_x$, $p_y$, and $p_z$-wave pair amplitudes as a function of $\beta$ in Fig.\,\ref{figp}. Here we fix all parameter values to that of Fig.\,\ref{figBeta} for easy comparison, and we also focus on a bulk layer ($n_{\text w}=19$) of the WSM in order to capture the bulk proximity effect. 

We first observe that the pair amplitudes of both the intra-orbital $p_x$ and $p_z$-wave pair amplitudes are suppressed by almost two orders of magnitude compared to the $s$-wave amplitudes, and they are also not surviving for larger $\beta$. For the inter-orbital components, the $p_x$ and $p_z$-wave components are also smaller by an order of magnitude compared with the $s$-wave amplitudes. Here we find that for the equal-spin triplet in particular, both ($\uparrow\uparrow$ and $\downarrow\downarrow$), the pair amplitude is growing with the increase of $\beta$. This increasing behavior of the inter-orbital $p_x$ and $p_z$-wave amplitudes is very similar to that of the $s$-wave inter-orbital pairing, but with different spin-structure and lower magnitude. 

Finally we concentrate on the $p_y$-wave pair amplitudes, which are, in contrast to the $p_x$ and $p_z$-wave components, actually comparable to or sometimes even larger than the $s$-wave amplitudes.
In particular, the spin-singlet and the mixed-spin triplet, both intra- and inter-orbital, $p_y$-wave pair amplitudes are very similar in size to the $s$-wave inter-orbital amplitudes with the same spin structure.  However, the equal-spin $p_y$-wave pair amplitudes are all small and in fact completely zero in the middlemost layer ($n_{\text w} = 21$). To summarize, there are thus also $p_y$-wave pair amplitudes, of both intra- and inter-orbital character, present in proximity-coupled magnetic WSMs that are not sensitive to any chirality blockade, in addition to the inter-orbital $s$-wave components.We give a comprehensive summary of the behaviors of all $p$-wave pair amplitudes in Table\,\ref{tab} with a comparison of their magnitudes to the corresponding $s$-wave pair amplitudes with the same spin structure.

\section{Josephson current}\label{JJ}
Our results in the previous section show that there are finite, and even large, pair amplitudes in magnetic WSMs proximity-coupled to conventional superconductors. These amplitudes have either $s$-wave inter-orbital symmetry, \ie connecting electrons between different orbitals, or $p$-wave spatial symmetry. When it comes to experimentally measurable quantities, however, the pair amplitude is not a good quantity as it is usually not directly accessible. The Josephson current between two phase-biased SCs is however one of the most used probes for establishing a superconducting proximity effect. 

In this section, we analyze the Josephson current in our LSC-WSM-RSC junction in order to capture the signatures of the proximity-induced pairing in an experimentally measurable quantity. More specifically, we calculate the supercurrent density using the free energy of the LSC-WSM-RSC Hamiltonian as
\bea
J(\varphi)=-\frac{e}{\hbar}\sum_{\tau,\sigma} \sum_{k_y,k_z} \frac{dE(\varphi)}{d\varphi}, 
\label{J_formula}
\eea
where $E$ are the energy levels of the whole system and we limit the summation to the already discussed $k_{y(z)}$ window (limited by $\pm k_c$). We set $e=\hbar=1$ here and thus express $J$ in natural units. 

\begin{figure}[htb]
\includegraphics[scale=0.65]{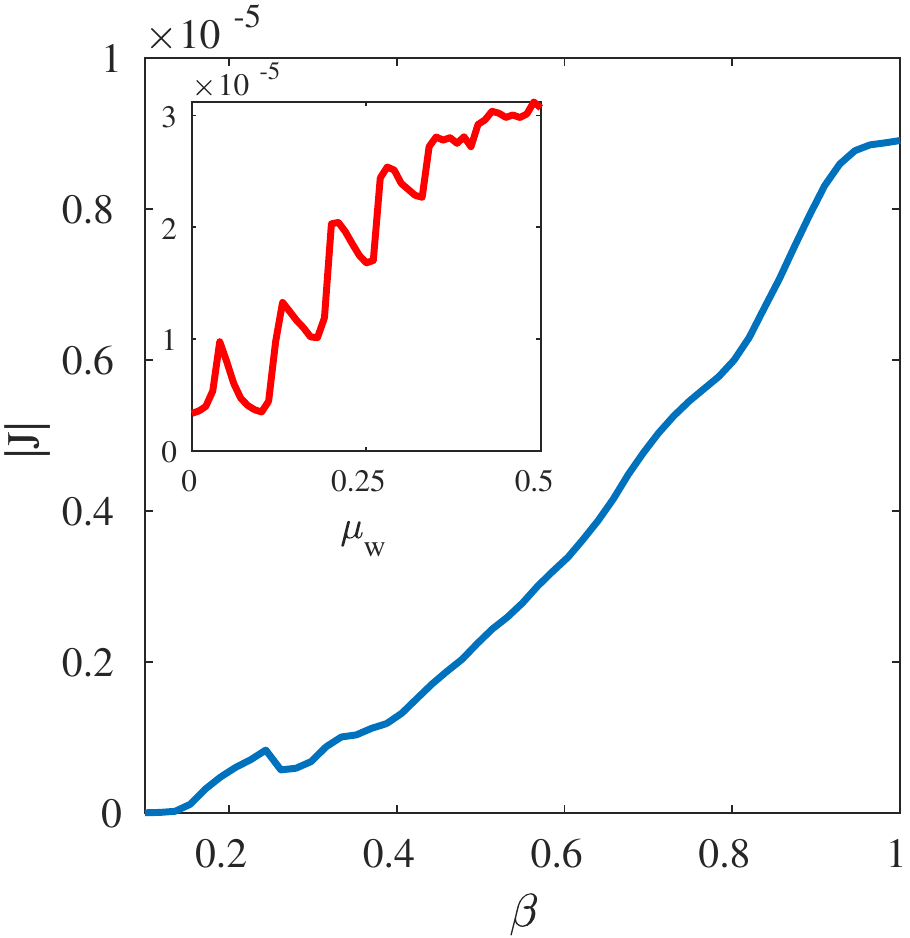}
\caption{Maximum Josephson current achieved at $\Delta \varphi=\pi/2$ as a function of $\beta$. Inset: Maximum Josephson current at $\Delta \varphi=\pi/2$ as a function of the chemical potential of the WSM for $\beta=0.6$. Other parameter values are same as in Fig.\,\ref{figBeta}.}
\label{figJ}
\end{figure}

We are mainly concerned with the maximum current, achieved for $\Delta\varphi = \pi/2$ or very close to it, which we then plot in Fig.\,\ref{figJ} as a function of $\beta$, keeping the other parameters the same as in Fig.\,\ref{figBeta}. As seen, the maximum current is almost always monotonically increasing with $\beta$. In the light of the chirality blockade, this increase is a completely unexpected behavior. Still, we can fully attribute this current to the bulk proximity effect as follows:
Current mediated by non-bulk effects must be carried by overlapping surface Fermi arcs, each with the penetration length $\xi_c\sim v_F/\beta$ into the bulk of WSM \cite{Beenakker2017, faraei2019}. Such surface contributions to the Josephson current thus decay exponentially as $exp(-L/\xi_c)$\,\cite{Beenakker2017}, and then necessarily give a decreasing current for increasing $\beta$. But this is completely opposite to the result in Fig.\,\ref{figJ}, and therefore the current in Fig.\,\ref{figJ} cannot have a surface origin.
Moreover, we choose to study long Josephson junctions with a junction length much larger than the superconducting coherence length $\xi \sim 10 a$ , such that the current, if only carried by surface states, would be very limited for all values of $\beta$. Finally, in the calculation of $J$ we have again limited the summation in $k$-space to an interval $\pm k_c$ around each Weyl node, thus automatically severely limiting the surface arc contribution, especially for the larger Weyl separations given by increasing $\beta$. We can thus definitively conclude that the current in Fig.\,\ref{figJ}, and especially its growth with increasing $\beta$, cannot be attributed to Fermi arc contributions but must necessarily be a WSM bulk property.

Having established that the increasing Josephson current in Fig.\,\ref{figJ} is a WSM bulk property, we must attribute the current to the finite $s$-wave inter-orbital $p_y$-wave intra- and inter-orbital pair amplitudes inside the WSM, which both grow with increasing $\beta$, exactly as the total current. 
In particular, for $s$-wave pairing it is only the spin-singlet and mixed-spin triplet inter-orbital pair amplitudes that survives at larger $\beta$ and can therefore also be the only components responsible for a finite Josephson current. In clean systems, such as the system we study, current contributions are also present from the $p_y$-wave pairing. As a consequence, the Josephson current in longer WSM junctions become a probe of the finite inter-orbital pairing.

Finally, we also check whether our results are sensitive to fine-tuning the chemical potential to the Weyl nodes by plotting the maximum current as a function of $\mu_{\text {W}}$ in the inset of Fig.\,\ref{figJ}. We here choose $\beta =0.6$ (the same choice as in Fig.\,\ref{figt}) for which a very clear chirality blockade of the intra-orbital $s$-wave pairing is achieved. From the behavior of $J$ we conclude that it is not necessary to fine-tune the chemical potential of the WSM to achieve a large Josephson current. Generally, we also find a larger current for larger doping, as expected. In reality the pair amplitude can also be slightly sensitive to the magnetization of the Weyl material through the inverse proximity effect. However, we find a very similar current even when we use a smaller $\Delta$, and we thus conclude that the inverse proximity effect has a negligible influence on our results.

\section{Concluding remarks}
\label{conclu}

To summarize, we have explored proximity-induced superconductivity in a Josephson junction formed by a magnetic WSM using conventional $s$-wave spin-singlet superconductors. We have found that conventional $s$-wave intra-orbital spin-singlet pairing cannot be proximity-induced in the WSM, as it is incompatible with the opposite chirality of the two Weyl nodes in a magnetic WSM. 
This finding is in full agreement with the previously predicted phenomenon of chirality blockade of Andreev reflection \cite{Beenakker2017}. We have also revealed a similar blockade for the $s$-wave intra-orbital mixed-spin triplet odd-$\omega$ pairing. Moreover, the $s$-wave intra-orbital equal-spin triplet odd-$\omega$ pairing also vanishes in the bulk layers due to the characteristic spin structure of the WSM. As a consequence, we find that no $s$-wave intra-orbital pairing can be proximity-induced into a magnetic WSM.

Now, we still find a large proximity effect well into the bulk of the magnetic WSM. This proximity effect is due to large induced $s$-wave pair amplitudes with an inter-orbital pair symmetry, existing in both the even- and odd-$\omega$ channels. The inter-orbital character removes the possibility of chirality blockade of Andreev reflection in the junction. We also find large $p_y$-wave pair amplitudes, with both intra- and inter-orbital, as well as mixed-spin triplet and spin-singlet, symmetries. While $p$-wave amplitudes are most often assumed to be disorder sensitive, we note that inter-orbital $s$-wave pairing has been found to be robust against disorder in systems with strong spin-orbit coupling, such as WSMs \cite{Michaeli&Fu12}. Thus, we expect a large proximity effect stemming from inter-orbital pairing, even in realistic magnetic WSM materials which contain a finite amount of disorder. In order to measure this proximity effect we have calculated the Josephson current and found that it increases with increasing magnetic ordering. Such Josephson current behavior is completely incompatible with surface contributions, but instead in agreement with the behavior of the inter-orbital $s$-wave and $p_y$-wave pair amplitudes. 
We therefore conclude that magnetic WSM Josephson junctions using conventional superconducting leads carry a finite Josephson current due to unconventional inter-orbital and $p$-wave pair amplitudes and that, in particular, the specific chirality of the Weyl nodes are not blocking a superconducting proximity effect.

Before ending we offer a few remarks on how our results are related to previous works. In Ref.~[\onlinecite{Beenakker2017}], Bovenzi \etal established a chirality blockade for the Andreev reflection but did so by only considering intra-orbital pairing. We use a very similar setup for the junction but explicitly study all possible pair amplitudes, including induced pairing in inter-orbital, spin-triplet, and odd-frequency channels. Our results not only confirm the chirality blockade for intra-orbital spin-singlet $s$-wave pairing but also extends the discussion of vanishing pairing to all intra-orbital spin-triplet symmetries. However, most importantly, our results show that the chirality blockade is not preventing inter-orbital $s$-wave pairing with both even- and odd-$\omega$ symmetry, as well as intra- and inter-orbital $p$-wave pairing in clean systems, and that these give rise to a bulk Josephson effect. In particular, we can firmly discard surface contributions to the Josephson effect, even though they have been shown to be present in longer junctions due to the chirality blockade\,\cite{shvetsov2019lateral}.  
Finite intra- and inter-orbital pairing have also been discussed in Ref.\,[\onlinecite{KhannaProximity}] for WSM-SC junctions. However, they found that no proximity-induced pair amplitude changes with the variation of the magnetization $\beta$ and that no chirality blockade exists for any pairing. This is completely at odds with our results and the difference can be explained by their usage of a pseudoscalar pair potential for the leads instead of using conventional $s$-wave SCs, as explained in Ref.\,[\onlinecite{Beenakker2017}]. 

\begin{acknowledgements}
We thank M.~Mashkoori and C.~Triola for useful discussions. We acknowledge the financial support from the Swedish Research Council (Vetenskapsr\aa det Grant No.~2018-03488), the Knut and Alice Wallenberg Foundation, and the European Research Council (ERC) under the European Unions Horizon 2020 research and innovation programme (ERC-2017-StG-757553). P.D. thanks Arijit Saha, Arijit Kundu, and Sumathi Rao for introducing her to WSMs.
\end{acknowledgements}

\begin{appendix}
\section{Lattice Hamiltonians}\label{apn1}
In this Appendix, we present the details of the lattice Hamiltonians used for the results in the main text. 
We discretize the WSM Hamiltonian in Eq.\,(\ref{Ham_wsm}) by taking the inverse Fourier transformation along the $x$-axis as
\bea
\bm{H}_{\text{\tiny WSM}} (k_y,k_z)&=&\sum\limits_r[ \bm{c}_{r,k_y,k_z}^{\dagger}(\tau_z \sum\limits_{i=y,z} t_i \sigma_i \sin{(k_i a)}+\beta \tau_0 \sigma_z
\nonumber \\
&&\left. +(m_0+t_x^{\prime}) \tau_x \sigma_0-\mu_{\text{\tiny W}}\right)\bm{c}_{r,k_y,k_z}]\nonumber \\
&&+[\bm{c}_{r,k_y,k_z}^{\dagger}(-t_x^{\prime} \tau_x \sigma_0-i t_x \tau_z \sigma_x)\,\bm{c}_{r+\delta r,k_y,k_z}\nonumber \\
&&+\text{H.c.}].
\label{H_w}
 \eea
Here $\bm{c}^{\dagger}_{r,k_y,k_z}$ ($\bm{c}_{r,k_y,k_z}$) is the creation (annihilation) operator for electrons in layer $r$ in a quasi one-dimensional ($1$D) chain along the $x$-axis with a periodic momentum description in $k_y$ and $k_z$, while $r$$+$$\delta r$ represents the position of the nearest neighbor layers along the $x$ axis. 

Similar to the WSM, we also take the inverse Fourier transformation along the $x$-axis in the SCs described by $\bm{H}^0_{\text{\tiny L(R)SC}}(\bm{k})$ in Eq.\,\eqref{hsc} and we arrive at
\bea 
\bm{H}^{\text 0}_{\text{L(R)SC}}(k_y,k_z)&=&\sum\limits_{r} \bm{b}^{\text{L(R)}\,\dagger}_{r,k_y,k_z} [-
\mu_{\text{SC}} + t_{\text{SC}}(2-\cos{(k_ya)} \nonumber \\
&&~~~~~~~~~~~~~~~~~~ -\cos{(k_z a)})]\ {\bm{b}}^{\text{L(R)}}_{r,k_y,k_z}\nonumber \\
&&+[ t_{\text{\tiny SC}}\, \bm{b}^{\text{L(R)}\,\dagger}_{r,k_y,k_z} \bm{b}^{\text{L(R)}}_{r+\delta, r,k_y,k_z} + \text{H.c.}],
\label{H_sc}
\eea
where $\bm{b}^{\text{L(R)}\dagger}_{r,k_y,k_z}$ ($\bm{b}^{\text{L(R)}}_{r,k_y,k_z}$) is the creation (annihilation) operator for electrons inside the LSC and RSC, respectively.

The tunneling between the WSM and each SC lead is described by a Hamiltonian given by\,\cite{Burkov,Beenakker2017}
\bea
\bm{H}_{\text T}(k_y,k_z)&=& t_{\text{\tiny W-SC}}\, (\bm{c}_{1,k_y,k_z}^{\dagger} \bm{b}^{\text{L}\,\dagger}_{N_{\text{SC}},k_y,k_z} \nonumber \\
&&+\bm{c}_{N_{\text WSM},k_y,k_z}^{\dagger} \bm{b}_{1,k_y,k_z}^{\text{R}} + \text{H.c.}),
\eea
where $t_{\text{\tiny W-SC}}$ is the coupling matrix between the WSM and each SC lead. It can be expressed as
\beq
t_{\text{\tiny W-SC}}=diag(t^a_{\text{\tiny W-SC}},t^{b}_{\text{\tiny W-SC}})
\eeq
where $t^{a\,(b)}_{\text{\tiny{W-SC}}}$ corresponds to the coupling for $a (b)$ orbital as mentioned in the main text. Here, we couple the last layer ($n_{L}=N_{\text{\tiny SC}}$) of the left SC lead to the first ($n_{\text w}=1$) layer of the WSM and similarly, the first layer ($n_R=1$) of the right lead to the last layer ($n_{\text{\tiny WSM}}=N_{\text{\tiny WSM}}$) of the WSM.
 
\end{appendix}
\bibliography{bibfile}

\end{document}